\documentclass[aps,prb,twocolumn,superscriptaddress,showpacs]{revtex4-1}
\usepackage{graphicx}
\usepackage{amsmath,amssymb,amsfonts}
\usepackage[hyperindex,breaklinks]{hyperref}
\usepackage{bm}

\newcommand{\Ham}{\mathcal{H}}
\newcommand{\bea}{\begin{eqnarray}}
\newcommand{\eea}{\end{eqnarray}}
\newcommand{\beq}{\begin{equation}}
\newcommand{\eeq}{\end{equation}}
\newcommand{\nn}{\nonumber}

\begin{document}

\title{Double magnetic resonance and spin anisotropy in Fe-based superconductors due to static and fluctuating antiferromagnetic orders}
\author{Weicheng Lv}
\affiliation{Department of Physics and Astronomy, University of Tennessee, Knoxville, Tennessee 37996, USA}
\affiliation{Materials Science and Technology Division, Oak Ridge National Laboratory, Oak Ridge, Tennessee 37831, USA}
\author{Adriana Moreo}
\affiliation{Department of Physics and Astronomy, University of Tennessee, Knoxville, Tennessee 37996, USA}
\affiliation{Materials Science and Technology Division, Oak Ridge National Laboratory, Oak Ridge, Tennessee 37831, USA}
\author{Elbio Dagotto}
\affiliation{Department of Physics and Astronomy, University of Tennessee, Knoxville, Tennessee 37996, USA}
\affiliation{Materials Science and Technology Division, Oak Ridge National Laboratory, Oak Ridge, Tennessee 37831, USA}

\date{\today}

\begin{abstract}
Motivated by recent neutron scattering experiments in Fe-based superconductors, we study how the magnetic resonance in the superconducting state
is affected by the simultaneous presence of either static or fluctuating magnetic orders using the random phase approximation. We find that for the underdoped materials with coexisting superconducting and antiferromagnetic orders, spin rotational symmetry is explicitly broken at the ordering momentum $\bm{Q}_1 = (\pi,0)$. Only the longitudinal susceptibility exhibits the resonance mode, whereas a spin-wave Goldstone mode develops in the transverse component. Meanwhile, at the frustrated momentum $\bm{Q}_2 = (0,\pi)$, the susceptibility becomes isotropic in spin space and the magnetic resonance exists for both components. Furthermore, the resonance energies at $\bm{Q}_1$ and $\bm{Q}_2$ have distinct scales, which provide a natural explanation for the recently observed double resonance peaks. In addition, we show that near optimal doping the existence of
strong magnetic fluctuations, which are modeled here via a Gaussian mode, can still induce the spin anisotropy in the magnetic susceptibility.
\end{abstract}


\maketitle


\section{Introduction}

A ubiquitous feature present in most unconventional superconductors is the development of a magnetic resonance mode\cite{Rossat-Mignod1991,Mook1993,Fong1995,Stock2008,Christianson2008,Lumsden2009,Chi2009,Yu2009a,Scalapino2012,Dai2012} as observed by inelastic neutron scattering (INS)
experiments below the superconducting
(SC) transition temperature, $T_c$. Despite some theoretical controversies,\cite{Bulut1993,Liu1995,Demler1995,Demler1998,Brinckmann1999,Tchernyshyov2001,
Lee2008a,Lee2008b,Lee2008b,Hao2009,Hao2009,
[{For a review, see }]Eschrig2006} the general consensus is that this resonance mode represents a spin exciton bound state in the particle-hole channel.
In such a framework, the appearance of the magnetic resonance is determined by the structure of the SC gap $\Delta(\bm{k})$ along
the Fermi surface via the BCS coherence factors. A sign difference between $\Delta(\bm{k})$ and $\Delta(\bm{k}+\bm{q})$ generates a resonance mode at momentum $\bm{q}$, with the resonance energy $\omega_\mathrm{res}$ below the onset of the particle-hole continuum, $\vert\Delta(\bm{k})\vert + \vert \Delta(\bm{k}+\bm{q}) \vert$. Usually the highest intensity of the resonance occurs at $\bm{q} = \bm{Q}$, the wave vector of the parent antiferromagnetic (AFM) order, which connects the Fermi surface points with the largest gap amplitudes. Located below the particle-hole continuum, and thus undamped by the quasiparticle excitations, the magnetic resonance is a $\delta$-function collective mode, which can appear as a sharp feature
in the INS experiments. Studying the detailed structure of the magnetic resonance provides important insights into the pairing symmetry and mechanism of unconventional superconductors.\cite{Maier2008,Korshunov2008,Maier2009,Zhang2010,Maiti2011b,Onari2011,Inosov2010,Park2010,
Castellan2011,Zhang2011a,Zhang2013a}

In principle, if the Cooper pair is a spin singlet and the magnetic interaction preserves the spin rotational symmetry, we expect that all three components of the dynamical spin susceptibility, $\chi({\bm{q},\omega})$, should be equal, i.e., the resonance should be isotropic in spin space.
However, recent polarized INS experiments have revealed different results for the Fe-based superconductors near optimal doping.\cite{Lipscombe2010,Steffens2013,Zhang2013,Luo2013a} In these materials, the out-of-plane susceptibility is consistently larger than the in-plane susceptibility below the resonance energy $\omega_\mathrm{res}$, whereas this spin anisotropy diminishes above $\omega_\mathrm{res}$. Furthermore, in overdoped materials, the magnetic susceptibility becomes fully isotropic within the current
experimental resolution.\cite{Liu2012a}

At first sight, the presence of the spin anisotropy may appear closely related to the proximity to the AFM order.
Because the ordered moments point along the in-plane $a$ direction,\cite{Cruz2008} spin rotational symmetry
is explicitly broken when the long-range AFM order sets in. Therefore, for the underdoped materials
with coexisting SC and AFM orders, the magnetic susceptibility should naturally develop an anisotropy in spin space.
However, polarized INS experiments have not been performed in this doping range yet. We may further conjecture
that with increasing doping, for the optimally doped materials at the boundary of the AFM phase, strong magnetic fluctuations
with a preferable spin direction could still induce an anisotropy in the spin susceptibility as a remnant
of the static long-range order. Finally, for the overdoped materials where the effect of the AFM fluctuations
becomes very weak, the magnetic resonance should display a spin rotational symmetry.

In this paper, we will study a three-band model using a generalized random phase approximation (RPA), and
investigate how the presence of static and fluctuating AFM orders can affect the spin anisotropy of the
magnetic resonance mode in Fe-based superconductors. In Sec.~\ref{sec:stat}, we first consider the spin susceptibility
in the underdoped materials in the presence of coexisting SC and AFM orders. Our results indicate that at the AFM order
momentum $\bm{Q}_1 = (\pi,0)$, the magnetic resonance only emerges in the longitudinal susceptibility,
whereas the transverse component exhibits a spin-wave Goldstone mode. By contrast, at the frustrated
momentum $\bm{Q}_2 = (0,\pi)$, spin rotational symmetry is preserved, with the resonance appearing
in both the transverse and longitudinal susceptibilities. Furthermore, the magnetic resonance shows
distinct energy scales at $\bm{Q}_1$ and $\bm{Q}_2$, which suggests a possible explanation for the
double resonance features that have been recently observed.\cite{Zhang2013b} In Sec.~\ref{sec:fluc}, we turn
to the case of the optimally doped materials with only the paramagnetic (PM) spin order. Using
a Gaussian mode to account for the strong AFM fluctuations, we find that the spin anisotropy
stills persists and mainly occurs below the resonance energy, in agreement with the INS experimental
results. Finally, our conclusions are summarized in Sec.~\ref{sec:sum}.

\section{\label{sec:stat}static magnetic order}

\subsection{RPA formalism in the coexistence phase}

The underdoped phase with coexisting SC and AFM orders in Fe-based superconductors has been studied extensively in previous efforts.\cite{Parker2009,Vorontsov2010,Fernandes2010b,Fernandes2010a,Fernandes2010,Knolle2011,Maiti2012} 
In particular, the magnetic resonance in the SC-AFM coexistence phase has been investigated
before in detail in Ref.~\onlinecite{Knolle2011}. In this section, we reexamine this issue
in light of the recently observed double resonance peaks in the underdoped NaFeAs.\cite{Zhang2013b}

The standard three-band model used in this publication
is defined in the extended Brillouin zone with one Fe per unit cell,
and it consists of a hole pocket $c$ at $\Gamma = (0,0)$ and two electron
pockets $f_1$ and $f_2$ at $\bm{Q}_1 = X = (\pi,0)$ and $\bm{Q}_2 = Y = (0,\pi)$,
\bea
    \Ham_K & = & \sum_{\bm{k}, \mu} \left(\epsilon_{c}(\bm{k}) - \mu\right) c_{\bm{k}\mu}^\dagger c_{\bm{k}\mu}
    \nn \\
    & & \,
    + \sum_{n=1,2}\sum_{\bm{k}, \mu} \left(\epsilon_{f_n}(\bm{k}) - \mu\right) f_{n,\bm{k}\mu}^\dagger f_{n,\bm{k}\mu}.
\eea
Each band has a simple quadratic dispersion,
\bea
    \epsilon_{c}(\bm{k}) & = & \epsilon_{c,0} - \frac{k_x^2+k_y^2}{2m}, \\
    \epsilon_{f_{1(2)}}(\bm{k}+\bm{Q}_{1(2)}) & = & -\epsilon_{f,0} + \frac{k_x^2}{2m_{x(y)}} + \frac{k_y^2}{2m_{y(x)}}.
\eea
We choose the same set of parameters as in Ref.~\onlinecite{Fernandes2010b}, which leads to the experimentally observed circular hole pocket and elliptical electron pockets. The inclusion of additional hole pockets is straightforward and does not change our results qualitatively.
We further comment briefly that a more accurate description of the electronic structure in Fe-based superconductors should be formulated in the orbital representation.\cite{Graser2009,Daghofer2010,Nicholson2011a} The use of the band representation here, however, can simplify our calculation and provide a clear physical picture.

We first write down the term corresponding to the SC order
in a general form as follows,
\bea
    \Ham_{\Delta}& = & \sum_{\bm{k}} \Delta_c(\bm{k}) \left( c_{\bm{k}\uparrow}^\dagger c_{-\bm{k}\downarrow}^\dagger + h.c. \right) \nn \\
    & & \, + \sum_{n=1,2} \sum_{k} \Delta_{f_n}(\bm{k}) \left( f_{n,\bm{k}\uparrow}^\dagger f_{n,-\bm{k}\downarrow}^\dagger + h.c. \right),
\eea
where $\Delta_c(\bm{k})$ and $\Delta_{f_{n}}(\bm{k})$ are the SC order parameters on the hole and electron bands, respectively. We simply take the $s_\pm$ gap structure and ignore any angular dependence, $\Delta_c(\bm{k}) = - \Delta_{f_{n}}(\bm{k}) = \Delta$. In principle, the gap amplitude $\Delta$ can be calculated from some microscopic theory,\cite{Vorontsov2010,Fernandes2010b,Fernandes2010a,Fernandes2010,Knolle2011} but for the purposes of simplicity,
we will treat this amplitude as a phenomenological parameter.

Then, we consider the magnetic interaction between the hole and electron bands,
\bea
    \Ham_\mathrm{AFM} & = & -\frac{2J}{N} \sum_{n=1,2} \sum_{\bm{k},\mu\nu} c_{\bm{k}\mu}^\dagger \frac{\bm{\sigma}_{\mu\nu}}{2} f_{n,\bm{k}+\bm{Q}_n, \nu} \nn \\
    & & \, \cdot
    \sum_{\bm{k}^\prime,\mu^\prime \nu^\prime} f_{n,\bm{k}^\prime+\bm{Q}_n, \mu^\prime}^\dagger \frac{\bm{\sigma}_{\mu^\prime \nu^\prime}}{2} c_{\bm{k}^\prime \nu^\prime},
\label{eq:HAFM}
\eea
where $N$ is the number of sites and $\bm{\sigma}_{\mu\nu}$ is the Pauli matrix. When the interaction strength $J$ is sufficiently large, a long-range AFM order sets in, with the ordering wave vector being either $\bm{Q}_1 = (\pi,0)$ or $\bm{Q}_2 = (0,\pi)$.\cite{Eremin2010,Brydon2011a,Fernandes2012} Without loss of generality, we assume that the AFM order occurs at $\bm{Q}_1$, involving the hole band $c$ and the electron band $f_1$. We will call $\bm{Q}_2$ the frustrated momentum, where the AFM order fails to develop. By projecting the ordered moment along the spin $z$ direction, we can write down the AFM order term in the mean-field form as,
\bea
    \Ham_{M} = M \sum_{\bm{k}, \mu} \mu \left(c_{\bm{k}\mu}^\dagger f_{1,\bm{k}+\bm{Q}_1,\mu} + h.c. \right),
\eea
where $\mu = \pm 1$ for up and down spins, respectively. The AFM order parameter $M$ can be calculated self-consistently from standard mean-field theory via,
\bea
    M = - \frac{J}{2N} \sum_{\bm{k}, \mu} \mu \left\langle c_{\bm{k}\mu}^\dagger f_{1,\bm{k}+\bm{Q}_1,\mu} \right\rangle.
\eea
This self-consistency procedure is necessary to obtain spin waves in the RPA formalism.\cite{Schrieffer1989,Chubukov1992,Brydon2009,Knolle2010,Kaneshita2010,Knolle2011a}

In summary, we have a quadratic Hamiltonian involving the SC and AFM order parameters $\Delta$ and $M$,
\bea
    \Ham [\Delta, M ] = \Ham_K + \Ham_\Delta + \Ham_M.
\label{eq:Ham}
\eea
Now we are ready to calculate the generalized dynamical spin susceptibility,
\bea
    \chi^{ij}\left(\bm{q},\bm{q}^\prime,i\omega \right) = \int_0^\beta \mathrm{d}\tau \left\langle T_\tau S^i(\bm{q},\tau)  S^j(-\bm{q}^\prime,0) \right\rangle e^{i\omega\tau},
\eea
where the spin operator is defined as,
\bea
    \bm{S}(\bm{q},\tau)
    & = & \sum_{a,b=c,f_1,f_2}  \frac{1}{\sqrt{N}} \sum_{\bm{k}} a_{\bm{k}\mu}^\dagger(\tau) \frac{\bm{\sigma}_{\mu\nu}}{2} b_{\bm{k}+\bm{q},\nu}(\tau) \nn \\
    & = & \sum_{a,b=c,f_1,f_2} \bm{S}_{ab}(q,\tau).
\eea
The total spin susceptibility should include all the combinations among $c$, $f_1$, and $f_2$,\cite{Brydon2009}
\bea
    \chi^{ij}(\bm{q},\bm{q}^\prime,i\omega)
    & = & \sum_{a,b} \sum_{a^\prime,b^\prime} \int_0^\beta \mathrm{d}\tau \left\langle T_\tau S_{ab}^i(\bm{q},\tau)  S^j_{a^\prime b^\prime}(-\bm{q}^\prime,0) \right\rangle e^{i\omega\tau} \nn \\
    & = & \sum_{a,b} \sum_{a^\prime,b^\prime} \chi^{ij}_{aba^\prime b^\prime}(\bm{q},\bm{q}^\prime,i\omega).
\eea

Note that in this multiband model, the AFM order arises from the interband
magnetic interaction $\Ham_\mathrm{AFM}$ (\ref{eq:HAFM}). Consequently, the spin operator $\bm{S}(\bm{q},\tau)$ is mainly contributed by the interband components, i.e.,
\bea
\bm{S}(\bm{q},\tau) = \left\{ \begin{array}{cc}
\bm{S}_{cf_1}(\bm{q},\tau) + \bm{S}_{f_1c}(\bm{q},\tau), & \bm{q}=\bm{Q}_1 \\
\bm{S}_{cf_2}(\bm{q},\tau) + \bm{S}_{f_2c}(\bm{q},\tau), & \bm{q}=\bm{Q}_2
\end{array}
\right. .
\eea
This approximation is valid\cite{Brydon2009} as long as $\bm{q}$ is close enough to $\bm{Q}_n$, whereas other components become more significant as $\bm{q}$ gets away from $\bm{Q}_n$. Since the focus of this work is the commensurate wave vector $\bm{q}=\bm{Q}_n$ and its
vicinity, we only take into account the contribution from the interband components to the spin susceptibility. Furthermore, the AFM order at $\bm{Q}_1=(\pi,0)$ breaks the translational symmetry, and thus the spin susceptibility $\chi^{ij}(\bm{q},\bm{q}^\prime,i\omega)$ should be nonzero at both $\bm{q}^\prime = \bm{q}$ and $\bm{q}^\prime = \bm{q}+\bm{Q}_1$. However, as only the interband components are of interest here, we only need to consider the case of $\bm{q}^\prime = \bm{q}$.

Therefore, the total spin susceptibility simply reads
\bea
\chi(\bm{q},i\omega)& = &\chi_{cfcf}(\bm{q},\bm{q},i\omega) + \chi_{cffc}(\bm{q},\bm{q},i\omega) \nn \\
 & & \, + \chi_{fccf}(\bm{q},\bm{q},i\omega) + \chi_{fcfc}(\bm{q},\bm{q},i\omega),
\eea
where $f$ stands for $f_1$ around $\bm{Q}_1 = (\pi,0)$ and $f_2$ around $\bm{Q}_2 = (0,\pi)$. The superscript $ij$ has been dropped for simplicity.
We can then write down $\chi(\bm{q},i\omega)$ in the form of a $2\times2$ matrix,
\bea
    \hat{\chi}(\bm{q},i\omega) = \left( \begin{array}{cc}
    \chi_{cfcf}(\bm{q},\bm{q},i\omega) & \chi_{cffc}(\bm{q},\bm{q},i\omega) \\
    \chi_{fccf}(\bm{q},\bm{q},i\omega) & \chi_{fcfc}(\bm{q},\bm{q},i\omega)
    \end{array} \right).
\eea
The RPA spin susceptibility can be easily calculated as,
\bea
    \hat{\chi}_\mathrm{RPA}(\bm{q},i\omega) = \frac{\hat{\chi}_0(\bm{q},i\omega)}
    {\hat{I} - \hat{U}\hat{\chi}_0(\bm{q},i\omega)},
\eea
where the interaction matrix is
\bea
    \hat{U} = \left( \begin{array}{cc}
    0 & J \\
    J & 0
    \end{array} \right),
\eea
and $\hat{\chi}_0(\bm{q},i\omega)$ is the bare susceptibility obtained from the non-interacting Hamiltonian of Eq.~(\ref{eq:Ham}). We do not consider other intraband and interband interactions,\cite{Maiti2010} which are not important for the purposes of this work.

The imaginary part of the susceptibility, $\mathrm{Im} \chi(\bm{q},\omega)$, which is measured by the INS experiments, can be evaluated using the standard analytical continuation, $i\omega \rightarrow \omega + i\delta$, with a finite damping factor $\delta = 0.003$. From now on, we adopt an implicit energy unit of $e$V unless noted otherwise. Our calculation is performed on a $1000 \times 1000$ momentum grid to achieve a sufficient energy resolution. Finally, we set the chemical potential $\mu=0$, so that the highest intensity of the resonance appears at commensurate wave vectors, which is consistent with the INS experimental observations.\cite{Christianson2009,Pratt2010,Wang2010a} Different choices of $\mu$ do not modify our results significantly unless the mismatch between the hole and electron pockets becomes so large that the resonance moves to an incommensurate momentum, as the previous study\cite{Knolle2011} has shown before.

\subsection{Results}

\begin{figure}
    \centering
    \includegraphics[width=8cm]{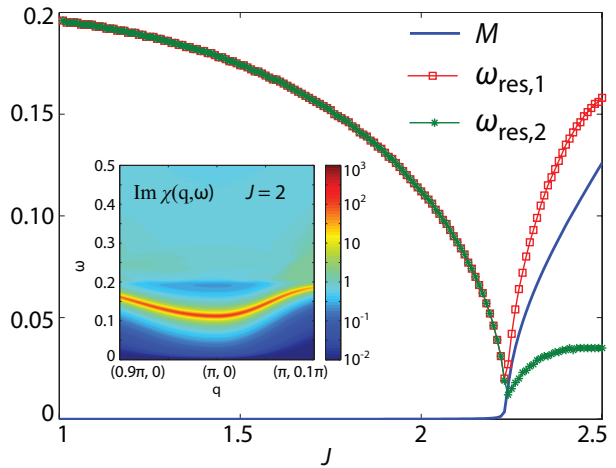}
    \caption{(Color online) The AFM order parameter $M$, the resonance energy $\omega_{\mathrm{res},1}$ at $\bm{Q}_1 = (\pi,0)$, and the resonance energy $\omega_{\mathrm{res},2}$ at $\bm{Q}_2 = (0,\pi)$, as functions of the AFM interaction strength $J$. We have set the SC order parameter to $\Delta = 0.1$. The inset shows the imaginary part of the spin susceptibility $\mathrm{Im} \chi(\bm{q},\omega)$ on a logarithmic scale in the PM phase with $J=2$, near $\bm{Q}_1 = (\pi,0)$ along the momentum cut $(0.9\pi,0) \rightarrow (\pi,0) \rightarrow (\pi,0.1\pi)$.}
    \label{fig:Mag_res}
\end{figure}

As a prerequisite, we need to determine the critical value of the magnetic interaction strength $J$, across which the system transitions from the PM to the AFM state. In Fig.~\ref{fig:Mag_res}, the AFM order parameter $M$ is plotted as a function of $J$, with the SC order
parameter fixed at $\Delta = 0.1$. From the figure, we can deduce
that the critical interaction strength is $J_c \approx 2.2$. It should be emphasized here that the value of $J_c$ depends on the magnitude of $\Delta$. Generally, a larger $\Delta$ leads to a higher $J_c$, which makes explicit the competition between the SC and AFM orders. In this section, we only consider the case of $\Delta=0.1$. But our main conclusion remains the same as long as $\Delta$ is chosen within the same order of magnitude.

Let us first calculate the spin susceptibility $\chi(\bm{q},\omega)$ in the PM phase, where we set $J =2 <J_c$. The inset of Fig.~\ref{fig:Mag_res} shows the imaginary part of the susceptibility, $\mathrm{Im} \chi(\bm{q},\omega)$, near $\bm{Q}_1 = (\pi,0)$. In the absence of the AFM order, the transverse and longitudinal components, $\mathrm{Im} \chi^{+-}$ and $\mathrm{Im} \chi^{zz}$, are identical, and the spin susceptibility around $\bm{Q}_2 = (0,\pi)$ can be simply obtained by a $\pi/2$ rotation. From this figure inset, we find that the resonance peak is centered at the commensurate momentum $\bm{Q}_1$ and exhibits an elliptical shape with elongation along the $x$ direction (longitudinal to $\bm{Q}_1$), which corresponds to the case of hole doping.\cite{Zhang2011a} This anisotropic momentum structure arises from the mismatch between the hole and electron pockets. If we modify the chemical potential $\mu$ to tune the system from hole doping to electron doping, the resonance peak will become elongated along the $y$ direction (transverse to $\bm{Q}_1$).\cite{Park2010}

Now we turn to the coexistence phase with both the SC and AFM orders. For that, we choose $J = 2.4 > J_c$, which gives rise to an AFM order parameter $M=0.0954$ for $\Delta = 0.1$. With this choice of $M$, the Fermi surface is still present when $\Delta$ vanishes, which indicates that the normal state is an AFM metal as observed in the underdoped regime. As $J$ keeps increasing, the Fermi surface will be eventually gapped out by a sufficiently large $M$, and the normal state becomes an AFM insulator. Although our results show no qualitative changes in this insulating phase, we will not consider this case because it is physically irrelevant for Fe-based superconductors.

\begin{figure}
    \centering
    \includegraphics[width=8cm]{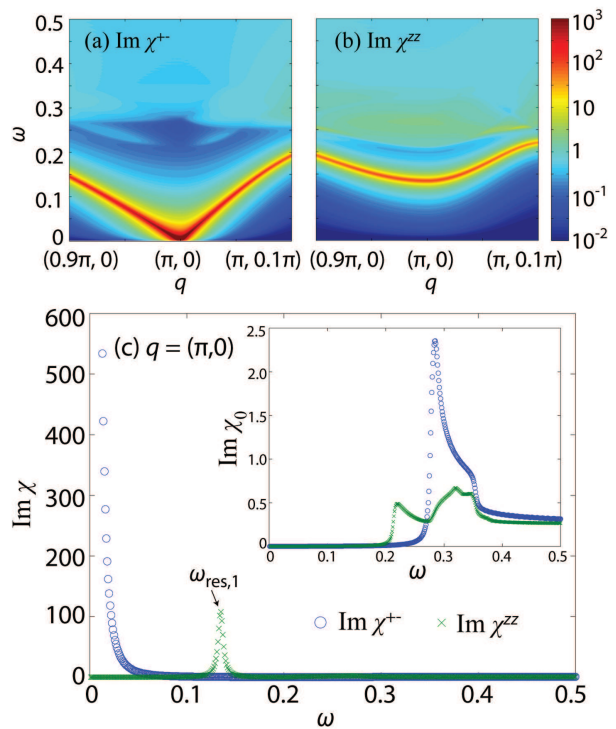}
    \caption{(Color online) (a), (b) The $(\bm{q},\omega)$ intensity map of the imaginary part of the spin susceptibility $\mathrm{Im}\chi(\bm{q},\omega)$ on a logarithmic scale, around the AFM order wave vector $\bm{Q}_1 = (\pi,0)$ along the momentum cut $(0.9\pi,0) \rightarrow (\pi,0) \rightarrow (\pi,0.1\pi)$. (a) The transverse component $\mathrm{Im}\chi^{+-}(\bm{q},\omega)$. (b) The longitudinal component $\mathrm{Im}\chi^{zz}(\bm{q},\omega)$. (c) The calculated $\mathrm{Im}\chi^{+-}(\bm{q},\omega)$ and $\mathrm{Im}\chi^{zz}(\bm{q},\omega)$ at momentum $\bm{q} = \bm{Q}_1 = (\pi,0)$, as functions of the frequency $\omega$. The inset shows the bare susceptibility $\mathrm{Im}\chi_0(\bm{q},\omega)$. The resonance energy at $\bm{Q}_1 = (\pi,0)$ is denoted by $\omega_{\mathrm{res},1}$. We have set $\Delta = 0.1$ and $J = 2.4$, which leads to $M = 0.0954$.}
    \label{fig:chi_Q1}
\end{figure}

As the spin rotational symmetry is explicitly broken by the presence of the AFM order here, we expect the transverse and longitudinal susceptibilities, $\chi^{+-}$ and $\chi^{zz}$, to be different. We first study the case where $\bm{q}$ is close to the AFM order momentum $\bm{Q}_1 = (\pi,0)$. In agreement with a previous theoretical study of the SC-AFM
coexistence phase,\cite{Knolle2011}
we find that a spin-wave Goldstone mode develops in the transverse susceptibility, $\mathrm{Im} \chi^{+-}(\bm{q},\omega)$ [Fig.~\ref{fig:chi_Q1}(a)]. This spin-wave mode shows anisotropic dispersions along the $x$ and $y$ directions, due to the ellipticity of the electron pockets.\cite{Knolle2010} On the other hand, the magnetic resonance only appears in the longitudinal susceptibility, $\mathrm{Im} \chi^{zz}(\bm{q},\omega)$, with a commensurate peak at $\bm{Q}_1 = (\pi,0)$  [Fig.~\ref{fig:chi_Q1}(b)]. Similar to the previous PM case, the elongation of the elliptical resonance peak is along the $x$ direction (longitudinal to the order momentum $\bm{Q}_1$). We should note that similar results, i.e., the transverse spin wave and longitudinal resonance,
have been obtained previously\cite{Rowe2012} in the context of the electron-doped cuprates,
which also exhibit a possible coexistence of SC and AFM orders. Finally, we plot both $\mathrm{Im}\chi^{+-}(\bm{q},\omega)$ and $\mathrm{Im}\chi^{zz}(\bm{q},\omega)$ as functions of $\omega$ at $\bm{q} = \bm{Q}_1$ in Fig.~\ref{fig:chi_Q1}(c). The peak at zero energy in $\mathrm{Im}\chi^{+-}$ represents the spin wave, whereas the magnetic resonance only occurs in $\mathrm{Im}\chi^{zz}$ with a peak energy $\omega_{\mathrm{res},1}$. It is noted that the spin-wave mode has a much larger spectral weight than the resonance. The inset of Fig.~\ref{fig:chi_Q1}(c) shows the bare susceptibility calculated with the noninteracting Hamiltonian of Eq.~(\ref{eq:Ham}). We see that the longitudinal component $\mathrm{Im} \chi_0^{zz}$ has a gap of $2\Delta = 0.2$, which suggests that it is only gapped by the SC order. By contrast, the transverse component $\mathrm{Im} \chi_0^{+-}$ is gapped by both the SC and AFM orders, with a gap amplitude approximately $2\sqrt{\Delta^2+M^2} \approx 0.28$. However, as mentioned previously, the AFM order parameter $M$ alone does not fully gap out the Fermi surface. Therefore, we do observe a small but finite value of $\mathrm{Im} \chi_0^{+-}$ below the estimated gap amplitude $2\sqrt{\Delta^2+M^2}$.

\begin{figure}
    \centering
    \includegraphics[width=8cm]{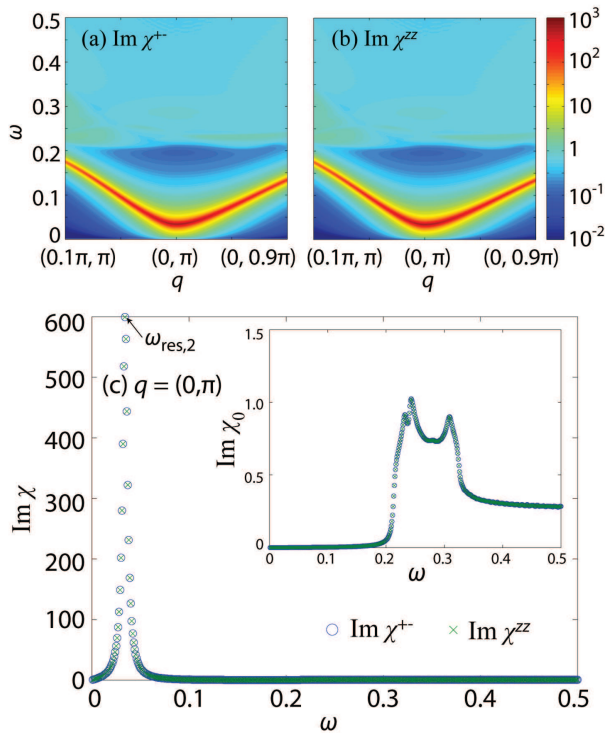}
    \caption{(Color online) (a), (b) The $(\bm{q},\omega)$ intensity map of the imaginary part of the spin susceptibility $\mathrm{Im}\chi(\bm{q},\omega)$ on a logarithmic scale, around $\bm{Q}_2 = (0,\pi)$ along the momentum cut $(0.1\pi,\pi) \rightarrow (0,\pi) \rightarrow (0,0.9\pi)$. (a) The transverse component $\mathrm{Im}\chi^{+-}(\bm{q},\omega)$. (b) The longitudinal component $\mathrm{Im}\chi^{zz}(\bm{q},\omega)$. (c) The calculated $\mathrm{Im}\chi^{+-}(\bm{q},\omega)$ and $\mathrm{Im}\chi^{zz}(\bm{q},\omega)$ at momentum $\bm{q} = \bm{Q}_2 = (0,\pi)$, as functions of the frequency $\omega$. The inset shows the bare susceptibility $\mathrm{Im}\chi_0(\bm{q},\omega)$. The resonance energy at $\bm{Q}_2 = (0,\pi)$ is denoted by $\omega_{\mathrm{res},2}$. We have set $\Delta = 0.1$ and $J = 2.4$, which leads to $M = 0.0954$.}
    \label{fig:chi_Q2}
\end{figure}

Experimentally, the INS measurement is usually performed on twinned samples and therefore it
has contributions from both types of twin domains. Namely, when the neutron scattering wave vector
is chosen to be the AFM order momentum in one type of the twin domains, the frustrated momentum in
the other type of twin domains will be measured simultaneously. So it is interesting to also study the behavior of the spin susceptibility near $\bm{Q}_2 = (0,\pi)$. Figs.~\ref{fig:chi_Q2}(a) and \ref{fig:chi_Q2}(b) contain $\mathrm{Im}\chi^{+-}(\bm{q},\omega)$ and $\mathrm{Im}\chi^{zz}(\bm{q},\omega)$, respectively, around the momentum $\bm{Q}_2$. Despite the fact that the hole pocket at $\Gamma = (0,0)$ is reconstructed by the AFM order parameter $M$, the spin susceptibility is completely isotropic in spin space, as the previous study\cite{Knolle2011} has suggested. The resonance occurs in both components of the spin susceptibility, $\mathrm{Im}\chi^{+-}$ and $\mathrm{Im}\chi^{zz}$. Again, this magnetic resonance is peaked at the commensurate momentum $\bm{Q}_2 = (0,\pi)$, with the elongation along the $y$ direction (longitudinal to $\bm{Q}_2$). In Fig.~\ref{fig:chi_Q2}(c), we further display the detailed energy dependence of both $\mathrm{Im}\chi^{+-}(\bm{q},\omega)$ and $\mathrm{Im}\chi^{zz}(\bm{q},\omega)$ at $\bm{q} = \bm{Q}_2$. Indeed, they are simply identical to each other, with a sharp resonance at $\omega_{\mathrm{res},2}$. The bare susceptibility in the inset of Fig.~\ref{fig:chi_Q2}(c) shows that both $\mathrm{Im}\chi_0^{+-}$ and $\mathrm{Im}\chi_0^{zz}$ are only gapped by the SC order, with a gap amplitude $2\Delta = 0.2$.

One interesting observation from our results is that in the coexistence phase the magnetic resonance modes exhibit distinct energy scales between the order momentum $\bm{Q}_1$ and the frustrated momentum $\bm{Q}_2$. According to Figs.~\ref{fig:chi_Q1}(c) and \ref{fig:chi_Q2}(c), the resonance energy $\omega_{\mathrm{res},1}$ at $\bm{Q}_1$ is much larger than $\omega_{\mathrm{res},2}$ at $\bm{Q}_2$. To better understand this, we plot $\omega_{\mathrm{res},1}$ and $\omega_{\mathrm{res},2}$ as functions of the AFM interaction strength $J$ in Fig.~\ref{fig:Mag_res}. In the PM phase with $J<J_c$, $\omega_{\mathrm{res},1}$ and $\omega_{\mathrm{res},2}$ are simply equal, approaching zero as $J$ increases to its critical value $J_c$. When the AFM order sets in for $J > J_c$, the magnetic resonance at the order momentum $\bm{Q}_1$ is split into a spin-wave mode at zero energy in the transverse component (not shown here) and a resonance mode at $\omega_{\mathrm{res},1}$ in the longitudinal component, whereas the resonance at $\bm{Q}_2$ remains isotropic in spin space with an energy $\omega_{\mathrm{res},2}$. As shown in Fig.~\ref{fig:Mag_res}, both $\omega_{\mathrm{res},1}$ and $\omega_{\mathrm{res},2}$ increase as the AFM order parameter $M$ grows, but with very different slopes. Qualitatively speaking, the resonance at $\bm{Q}_1$ is affected by both the SC and AFM orders, and thus has a higher energy $\omega_{\mathrm{res},1}$. By contrast, the AFM order has no impact on the resonance at $\bm{Q}_2$, except to reconstruct the hole pocket at $\Gamma = (0,0)$. Therefore, $\omega_{\mathrm{res},2}$ remains small in the coexistence phase.

In principle, as mentioned previously, for the INS experiments on twinned samples, the resonance modes
at both momenta, $\bm{Q}_1$ and $\bm{Q}_2$, will simultaneously contribute to the signal when the neutron scattering
wave vector equals either one or the other of the AFM order momenta. The question is whether the difference between $\omega_{\mathrm{res},1}$ and $\omega_{\mathrm{res},2}$ is large enough so that the two resonances can be clearly distinguished. Furthermore, in real materials, the SC gap amplitude has a strong angular dependence,\cite{Ge2013} which causes the resonance to obtain a finite width in energy instead of the $\delta$-functional form shown in our calculations. Consequently, in earlier INS experiments on the BaFe$_2$As$_2$ superconductors, only one single resonance with a very broad width was observed. But it was also found that the ratio of the resonance energy to the SC gap amplitude, $\omega_{\mathrm{res}}/\Delta$, was generally larger in the underdoped materials than in the optimally doped and overdoped materials.\cite{Inosov2011} From our theory, it is the higher resonance energy $\omega_{\mathrm{res},1}$ at $\bm{Q}_1$ that makes the ratio $\omega_{\mathrm{res}}/\Delta$ larger in the coexistence phase. Nevertheless, the predicted feature of the two resonance peaks has already been observed very recently in the underdoped NaFeAs superconductors.\cite{Zhang2013b} Our theory thus suggests a natural explanation that these two resonances actually come from two different momenta, $\bm{Q}_1$ and $\bm{Q}_2$. This scenario is distinct from other earlier theoretical proposals,\cite{Das2011,Yu2014} which rely on the difference in the SC gap amplitudes on different parts of the Fermi surface. Finally, we point out that the
interpretation of the INS measurements can be more complicated than discussed thus far
due to the presence of  the spin-wave mode [see Fig.~\ref{fig:chi_Q1}(a)],
which should also appear at finite energy in real systems. In order to clearly distinguish between
all these different collective modes discussed here, it would be important
to perform temperature-dependent INS experiments on detwinned samples.\cite{zhangdai}

Before ending this section, we need to emphasize that in reality, the N\'eel temperature $T_N$ is higher than the SC transition temperature $T_c$. Thus the SC order actually occurs on the Fermi surface that has already been reconstructed by the AFM order. So from an experimental point of view, one should start with a preexisting AFM order and calculate the SC gap amplitude on the reconstructed Fermi surface, as in Ref.~\onlinecite{Maiti2012}. However, as our intention here is to study how the presence of the magnetic order modifies the resonance mode, we have taken a different approach, i.e., the AFM order develops in a state that is already superconducting.

\section{\label{sec:fluc}fluctuating magnetic order}

\subsection{Formalism of the Gaussian fluctuations}
As the previous section has shown, for the underdoped materials with coexisting SC and AFM orders, the spin susceptibility $\chi(\bm{q},\omega)$ becomes anisotropic in spin space close to the AFM order momentum $\bm{Q}_1$. But from the recent INS experiments,\cite{Lipscombe2010,Steffens2013,Zhang2013,Luo2013a} this spin anisotropy still exists for the optimally doped materials where the long-range AFM order vanishes. In this section, we will generalize our previous work in Sec.~\ref{sec:stat} to the PM phase in the presence of strong AFM fluctuations.

In general, we can write down a Ginzburg-Landau(GL) theory for the SC and AFM order parameters $\Delta$ and $M$ as:\cite{Fernandes2010}
\bea
    \mathcal{F} \left( \Delta, M \right) & = & a_s \Delta^2 + b_s \Delta^4 +  a_m M^2 + b_m M^4 \nn \\
    & & \, + g\Delta^2 M^2,
\eea
where the last term with a positive coefficient $g$ represents the interaction between $\Delta$ and $M$. In principle, this GL free energy can be derived from a more fundamental microscopic theory,\cite{Fernandes2010} but here we simply take it
as an effective phenomenological model. By minimizing $\mathcal{F} \left( \Delta, M \right)$ with respect to $\Delta$ and $M$, we obtain the mean-field solutions $\langle \Delta \rangle$ and $\langle M \rangle$. In the coexistence phase of the SC and AFM orders considered previously in Sec.~\ref{sec:stat}, both $\langle \Delta \rangle$ and $\langle M \rangle$ are nonzero. For the optimally doped materials that are our focus in this section, we have $\langle \Delta \rangle \neq 0$ and $\langle M \rangle = 0$.
Of course, if we only use the mean-field values of these parameters, there exists no anisotropy in the spin susceptibility
as the spin rotational symmetry is strictly preserved.

Therefore, we need to go beyond the mean-field theory and allow for the AFM order to fluctuate,
i.e., $\langle M^2 \rangle \neq 0$. In this case, $\langle \Delta \rangle$ will depend on $M^2$,
\bea
    \langle \Delta \rangle = \left\{ \begin{array}{cc}
    \sqrt{\Delta_0^2 - \kappa M^2}, & \Delta_0^2 > \kappa M^2 \\
    0, & \Delta_0^2 < \kappa M^2
    \end{array} \right. ,
\label{eq:Delta}
\eea
where we have defined $\Delta_0 = \sqrt{-a_s/2b_s}$ and $\kappa = g/2b_s$. $\Delta_0$ is simply the mean-field value of the SC order parameter $\Delta$ in the absence of the AFM order parameter $M$. According to Eq.~(\ref{eq:Delta}), the AFM fluctuations suppress the SC order, and the SC order may even vanish if the fluctuation strength is sufficiently large. We can write down the partition function with a fluctuating AFM order parameter up to the leading order as follows,
\bea
    \mathcal{Z} & = & \int \mathcal{D}M \mathcal{D}c^\dagger \mathcal{D}c \mathcal{D}f^\dagger \mathcal{D}f \exp\left( -M^2/\eta^2 \right) \nn \\
    & & \, \exp\left(-\int_0^\beta \mathrm{d}\tau c^\dagger \partial_\tau c + f^\dagger \partial_\tau f + \Ham [ \Delta, M] \right),
\label{eq:par_fluc}
\eea
where $f$ is the shorthand notation for both $f_1$ and $f_2$. The Hamiltonian $\Ham[ \Delta, M]$ adopts the previous form of Eq.~(\ref{eq:Ham}), where the SC order parameter $\Delta$ takes the mean-field value as defined in Eq.~(\ref{eq:Delta}). The AFM fluctuations are approximated by a Gaussian mode, where $\eta$ controls the strength of the fluctuations. Then, the dynamical spin susceptibility in the presence of the AFM fluctuations can be calculated as,
\bea
    \chi(\bm{q},\omega) = \int \mathrm{d}M e^{-M^2/\eta^2} \chi(\bm{q},\omega) [\Delta, M],
\label{eq:chi_fluc}
\eea
where $\chi(\bm{q},\omega)[\Delta, M]$ is obtained using the RPA approach for the Hamiltonian $\Ham[\Delta, M]$ defined previously.

A similar Gaussian theory\cite{Lee2012b} has been proposed previously to account for the orbital fluctuations that precede the structural phase transition in Fe-based superconductors. Here we briefly comment about the validity of this approach for the AFM fluctuations. First of all, only the magnitude of the AFM order $M$ is allowed to fluctuate in our theory, whereas the order momentum is fixed at $\bm{Q}_1 = (\pi,0)$ and the order direction is locked along the spin $z$ direction. In principle, both of them should be fluctuating. Therefore, instead of an integral over a scalar field $M$ as shown in Eqs.~(\ref{eq:par_fluc}) and (\ref{eq:chi_fluc}), we need to integrate over an infinite number of vector fields $\bm{M}(\bm{q})$. Evaluating such an integral is numerically challenging. However, we expect that the component with $\bm{q} = \bm{Q}_1$ and $\bm{M}$ along the direction of the ordered moment has the largest weight and makes the dominant contribution. So our results should stay qualitatively the same even when these additional fluctuations are taken into account. Second, the SC and AFM order parameters, $\Delta$ and $M$, are not treated on an equal footing. Namely, we allow $M$ to fluctuate whereas $\Delta$ only takes its mean-field expectation value as in Eq.~(\ref{eq:Delta}). This approach is, however, reasonable considering that our purpose is to study how the AFM fluctuations affect the magnetic resonance in the SC state. The key assumption of our theory is that the AFM fluctuations have a preferable direction, which is supported by several recent INS measurements.\cite{Luo2013a,Qureshi2012,Song2013} Experimentally, it has been established that for both the parent compounds\cite{Qureshi2012,Song2013} and optimally doped materials,\cite{Luo2013a} the magnetic fluctuations are indeed anisotropic in spin space in both the PM and AFM phases. As these studies\cite{Luo2013a,Qureshi2012,Song2013} have pointed out, the spin anisotropy may come from the spin-orbit coupling. Actually a recent theoretical effort\cite{Korshunov2013}
has confirmed that the spin-orbit coupling is indeed capable of inducing the observed
spin anisotropy below the resonance energy. In this regard, our work in this section is not about
the microscopic origin of the spin anisotropy. Instead, we introduce the anisotropic AFM fluctuations
as a phenomenological input of our theory and show how they can affect the resonance modes in the SC state.

\subsection{Results}

From Sec.~\ref{sec:stat}, the magnetic susceptibility is isotropic in spin space at the frustrated momentum $\bm{Q}_2 = (0,\pi)$. So for the PM phase studied in this section, we expect that the AFM fluctuations at $\bm{Q}_1 = (\pi,0)$ can lead to the spin anisotropy only at the same momentum $\bm{Q}_1$,  but not at the conjugate momentum $\bm{Q}_2$. However, we know that $\bm{Q}_1$ and $\bm{Q}_2$ are simply equivalent in the PM phase where the long-range AFM order vanishes. (Note that we do not consider the possible nematic phase\cite{Fernandes2012,Hu2012d,Fernandes2012a,Lee2013} here.) Therefore, the AFM fluctuations should also occur at $\bm{Q}_2$ with the same fluctuation strength as those at $\bm{Q}_1$, and they are responsible for inducing the same spin anisotropy at $\bm{Q}_2$. In this section, we only calculate the spin susceptibility at $\bm{Q}_1$ by taking into account the AFM fluctuations at the same momentum $\bm{Q}_1$. The case at $\bm{Q}_2$ is simply identical by symmetry. Finally, we note that once the long-range AFM order sets in at $\bm{Q}_1 = (\pi,0)$, the AFM fluctuations at $\bm{Q}_2 = (0,\pi)$ will become strongly suppressed, and thus the spin rotational symmetry will be restored at $\bm{Q}_2$ as already shown in Sec.~\ref{sec:stat}.

\begin{figure}
    \centering
    \includegraphics[width=8cm]{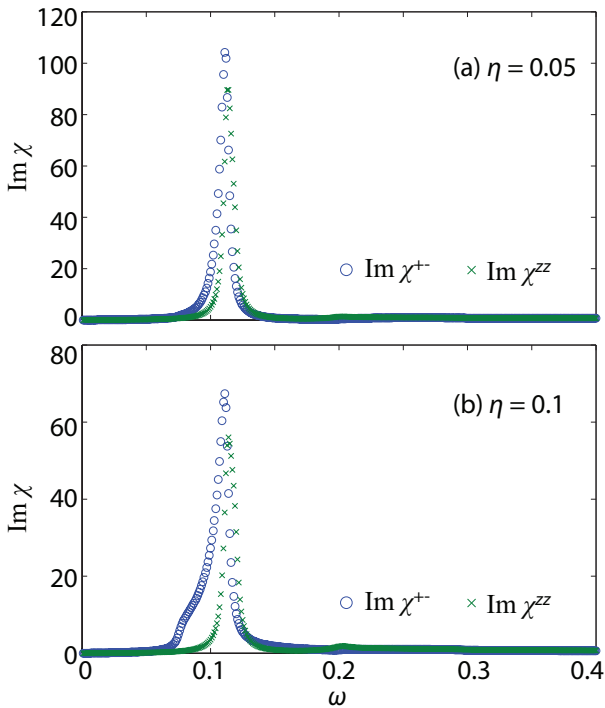}
    \caption{(Color online) The imaginary part of the transverse and longitudinal spin susceptibility, $\mathrm{Im} \chi^{+-}(\bm{q},\omega)$ and $\mathrm{Im} \chi^{zz}(\bm{q},\omega)$, as functions of the frequency $\omega$, at the momentum $\bm{q} = \bm{Q}_1 = (\pi,0)$. The AFM fluctuations at the same momentum are taken into account by a Gaussian mode, with the fluctuation strengths being (a) $\eta=0.05$ and (b) $\eta = 0.1$, respectively. We have set the other parameters as $\Delta_0 = 0.1$ and $\kappa = 1$. The AFM interaction $J$ is chosen appropriately for the PM phase, at $J=2<J_c$.}
    \label{fig:fluc}
\end{figure}

The calculated spin susceptibilities in the presence of the AFM fluctuations are plotted in Fig.~\ref{fig:fluc}.
As discussed previously, we only consider the commensurate momentum $\bm{Q}_1 = (\pi,0)$, where the resonance mode
has the highest intensity. The relevant parameters of our theory are chosen as $\Delta_0 = 0.1$ and $\kappa = 1$.
The AFM interaction strength is set at $J = 2 <J_c$, so that the system stays in the PM phase without any static
long-range magnetic order. We study two cases with different strengths of the AFM fluctuations, i.e., $\eta = 0.05$
in Fig.~\ref{fig:fluc}(a) and $\eta = 0.1$ in Fig.~\ref{fig:fluc}(b).

We find that as the fluctuation strength $\eta$ increases, the anisotropy between $\mathrm{Im} \chi^{+-}$ and $\mathrm{Im} \chi^{zz}$ also increases. Compared to the case without the AFM fluctuations at $\eta = 0$, as displayed in the inset of Fig.~\ref{fig:Mag_res}, the peak positions of these resonance modes stay almost the same. The spin anisotropy mainly occurs below the resonance energy, where the transverse component $\mathrm{Im} \chi^{+-}$ displays a larger weight than the longitudinal component $\mathrm{Im} \chi^{zz}$. By contrast, the magnetic resonances in both $\mathrm{Im} \chi^{+-}$ and $\mathrm{Im} \chi^{zz}$ exhibit a reduced height as the fluctuation strength $\eta$ increases, simply implying that the AFM fluctuations suppress the SC order. If $\eta$ further grows, the spin anisotropy will extend above the resonance energy (results not shown here), which is not observed in the INS experiments. Therefore, in reality, the AFM fluctuation strength should be smaller compared to the magnitude of the SC order parameter.
This is certainly a reasonable result, otherwise the SC order will be strongly suppressed near optimal doping.

Finally, we point out that it has been shown experimentally that actually all three components of the spin
susceptibility are different.\cite{Luo2013a,Song2013} But in our calculations, the two transverse components are
always the same because only one preferable direction, the spin $z$ direction, is selected for the AFM fluctuations.
Therefore, in this regard, our work should only be considered as a proof-of-principle study to show how the
magnetic resonance is modified by the anisotropic AFM fluctuations.

\section{\label{sec:sum}Summary}
To conclude, in this manuscript we have studied how the static and fluctuating AFM orders affect the
magnetic resonance mode in Fe-based superconductors. For the underdoped materials with coexisting SC
and AFM orders, spin anisotropy occurs around the AFM order momentum $\bm{Q}_1 = (\pi,0)$. The resonance
mode only exists in the longitudinal component $\mathrm{Im} \chi^{zz}$, whereas in the transverse component
$\mathrm{Im} \chi^{+-}$ is replaced by a spin-wave Goldstone mode. By contrast, at the frustrated momentum
$\bm{Q}_2 = (0,\pi)$, the spin rotational symmetry is preserved, with the resonance appearing in both
$\mathrm{Im} \chi^{+-}$ and $\mathrm{Im} \chi^{zz}$. Furthermore, we find that the resonance at $\bm{Q}_1$
occurs at a much higher energy than the resonance at $\bm{Q}_2$. These two resonance energy scales provide
a natural explanation for the double resonance peaks observed in the recent INS measurements on twinned
samples. Finally, we consider the optimally doped materials where the static long-range AFM order vanishes.
It is assumed that in this case strong AFM fluctuations still persist with a preferable order direction.
By modeling these fluctuations as a Gaussian mode, we show that the spin anisotropy in the magnetic
susceptibility occurs dominantly below the resonance energy, where the transverse component
$\mathrm{Im} \chi^{+-}$ is significantly larger than the longitudinal one $\mathrm{Im} \chi^{zz}$.
Overall, our work shows that the interplay between the SC and AFM order parameters can lead to interesting experimental
consequences for the magnetic resonances in Fe-based superconductors.

\begin{acknowledgments}
We would like to thank Pengcheng Dai and Chenglin Zhang for sharing of their experimental data and stimulating discussions, and Ilya Eremin and Thomas Maier for helpful comments about the manuscript.
This work was supported by the National Science Foundation Grant No.~DMR-1104386.
\end{acknowledgments}

\bibliography{C:/Dropbox/Research/bib/biblio}
\bibliographystyle{apsrev4-1}

\end{document}